\UseRawInputEncoding
\documentclass[prl,aps,twocolumn,superscriptaddress,amsmath,amssymb]{revtex4}
\usepackage{amsmath}
\usepackage[english]{babel}
\usepackage{graphicx}
\usepackage{bm}
\usepackage{color}

\newcommand{\ds}{\displaystyle}

\begin{document}

\title{Inverse Faraday Effect for Superconducting Condensates}

\author{S. V. Mironov}
\affiliation{Institute for Physics of Microstructures, Russian Academy of Sciences, 603950 Nizhny Novgorod, GSP-105, Russia}
\author{A. S. Mel'nikov}
\affiliation{Institute for Physics of Microstructures, Russian Academy of Sciences, 603950 Nizhny Novgorod, GSP-105, Russia}
\affiliation{University Bordeaux, LOMA UMR-CNRS 5798, F-33405 Talence Cedex, France}
\author{I. D. Tokman}
\affiliation{Institute for Physics of Microstructures, Russian Academy of Sciences, 603950 Nizhny Novgorod, GSP-105, Russia}
\author{V. Vadimov}
\affiliation{Institute for Physics of Microstructures, Russian Academy of Sciences, 603950 Nizhny Novgorod, GSP-105, Russia}
\affiliation{QCD Labs and MSP group, QTF Centre of Excellence, Department of Applied Physics, Aalto University, P.O. Box 15100, FI-00076 Aalto, Espoo, Finland}
\author{B. Lounis}
\affiliation{Institut d'Optique, LP2N UMR-CNRS 5298, F-33400 Talence, France}
\affiliation{University Bordeaux, LP2N, F-33400 Talence, France}
\author{A. I. Buzdin}
\affiliation{University Bordeaux, LOMA UMR-CNRS 5798, F-33405 Talence Cedex, France}
\affiliation{World-Class Research Center ``Digital biodesign and personalized healthcare'', Sechenov First Moscow State Medical University, Moscow 119991, Russia}

\date{\today}

\begin{abstract}
The Cooper pairs in superconducting condensates are shown to acquire a temperature - dependent dc magnetic moment under the effect of the circularly polarized electromagnetic radiation. The mechanisms of this inverse Faraday effect are investigated within the simplest version of the phenomenological dynamic theory for superfluids, namely, the time - dependent Ginzburg -- Landau (GL) model. The light induced magnetic moment is shown to be strongly affected by the nondissipative oscillatory contribution to the superconducting order parameter dynamics which appears due to the nonzero imaginary part of the GL relaxation time. The relevance of the latter quantity to the Hall effect in superconducting state allows to establish the connection between the direct and inverse Faraday phenomena.
\end{abstract}

\maketitle

The exploration of mechanisms allowing the generation and control of magnetic moment in solids solely by light has always been an attractive challenge for the condensed matter physics. The first systematic studies of the interplay between magnetism and optics are dated back to the works of Faraday, who discovered the rotation of the plane of the light polarization by a magnetic field. The inverse effect, namely, the generation of dc magnetic moment by the circularly polarized light (so-called inverse Faraday effect) was predicted lately by Pitaevskii \cite{Pitaevskii} and then observed in non-absorbing Eu$^{+3}$:CaF garnet \cite{Ziel}.  Currently, the interaction of the very short laser pulses with non-absorbing media is a rapidly expanding research topic of modern magnetism \cite{Kimel_1, Kimel_2}, and it has been clearly demonstrated that for the {\it insulating magnetic} materials the subpicosecond polarized laser pulses provide a tool for the magnetic moment manipulation at the femtosecond time scale \cite{Kimel_3}.

In contrast, the optical generation of the magnetic moment in conductive materials still remains challenging. There are several experimental hints supporting the observation of the inverse Faraday effect (IFE) in GdFeCo magnetic metallic amorphous alloy \cite{Kimel_4}, however, up to now the evidence of the IFE in {\it non-magnetic metals} has been reported only in a few publications (see, e.g., \cite{Sheldon}). Theoretical works are also rather scarce in this domain: the estimates of the IFE in metal plasma have been suggested in \cite{Hertel1, Hertel2, Battiato}, a very simplified ``harmonic atom'' model has been studied in \cite{Tokman}, the case of semiconductors has been approached in \cite{Quinteiro} and persistent currents appearing in ballistic nanorings due to IFE were treated in \cite{Koshelev1, Kibis}. Despite the strong differences in the electronic band structure of these materials the physics beyond the IFE in conductive finite-size samples is rather generic. The electric field ${\bf E}$ of the circularly polarized wave propagating perpendicular to the sample surface induces the excess charge density $\rho$ at the sample edges. The spatial rotation of the vector ${\bf E}$ produces the corresponding in-phase motion of the charge $\rho$ with the velocity ${\bf v}$ along the sample boundary which results in the nonzero time-averaged edge current ${\bf j}=\left<\rho{\bf v}\right>$, and, thus, emergence of dc magnetic moment.

During the past decade it became possible to study the IFE in artificial superfluid systems. In particular, it was experimentally demonstrated that illumination of the toroidal atomic Bose-Einstein condensate by the twisted light carrying non-zero angular momentum produces d.c. persistent supercurrents \cite{Ryu, Beattie}. The superfluids seem to provide a promising playground to study the IFE since the optically-induced supercurrents should survive even after  switching off the light. One can expect that similar light-stimulated persistent currents should emerge also in conventional solid-state superconductors, which may open the way for the ultra-fast optical control of magnetic states in the devices of superconducting spintronics \cite{Linder}. However, at the moment both the theory and experiment dealing with the IFE in superconductors is lacking.

\begin{figure}[t!]
\includegraphics[width=0.25\textwidth]
{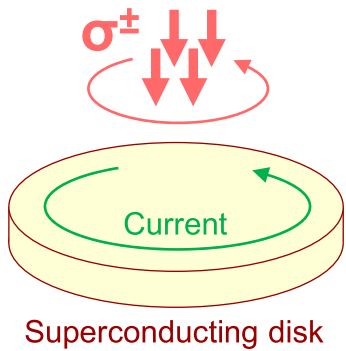}
\caption{Sketch of the thin superconducting disk radiated by the circularly polarized electromagnetic wave with two different polarizations ($\sigma^\pm$). The radiation-induced edge currents generate the magnetic moment of the disk (inverse Faraday effect).} \label{Fig1}
\end{figure}

It is the goal of the present Letter to suggest a theoretical description of the IFE in superconductors which may help to elaborate an appropriate experimental setup needed to observe the light stimulated magnetic states in superconducting systems (see the exemplary setup in Fig.~\ref{Fig1}). To elucidate the key ideas of our work we start from the qualitative consideration of the specific features of the IFE in superconductors.  First obvious difference between the IFE in normal metals and superconductors is based on the difference in dc magnetic response. Indeed, the magnetic response of the non-ferromagnetic normal metal to the dc magnetic moment $M_0$ generated by the electromagnetic radiation is determined by the Pauli and Landau terms in the susceptibility and both these terms are known to be extremely small \cite{Landau}. In opposite, the magnetic moment in superconductor should be screened by the Meissner supercurrents. The  screening can be partial or full depending on the sample geometry and the ratio of the its size to the London penetration depth. As a result, the total dc magnetic moment in the superconducting state can be partially or fully suppressed by the screening currents. This compensation will be broken when we increase the induced magnetic moment due to the vortex entry into the sample. To sum up, the overall picture is very similar to the one in the ferromagnetic superconductors \cite{Buzdin_rev} if we take account only of the Meissner screening phenomenon. These screening effects do not relate to the physical mechanisms of the $M_0$ formation though they may be important for the particular experimental setup and system configuration.

Focusing now on the physics of formation of the primary magnetic moment $M_0$ one can take a general relation between the current density and vector potential
${\bf j}(t)=\hat Q(N,\Delta){\bf A}(t)$, where the integral operator $\hat Q$ depends on the electron concentration $N$, superconducting gap $\Delta$ and also accounts for the relevant time dispersion. According to the Ref.~\cite{Hertel1} the IFE can be obtained if we consider the second order nonlinear corrections in the field of the electromagnetic wave in the above material relation. Thus, we need the corrections to the operator $\hat Q$ linear in the electric field ${\bf E}$. These corrections may originate from the
deviations in the local concentration $\delta N\propto {\rm div} {\bf E}$ from its equilibrium value $N_0$. The simplest expansion for the current density which assumes the local in time dependence on the deviation $\delta N (t)$ reads:
${\bf j}(t)=\hat Q_0 {\bf A}(t)+\left[\delta N (t)/N_0\right]\hat Q_1 {\bf A}(t)$.
In the frequency representation (for the $e^{-i\omega t}$ processes) the operator $\hat Q_1$ can be written as follows:
\begin{equation}
Q_{1}(\omega) = N_0\left(\frac{\partial Q_0(\omega)}{\partial N}+ \frac{\partial  Q_0(\omega)}{\partial \Delta}\frac{\partial \Delta}{\partial \mu} \frac{\partial \mu}{\partial N} \right)\ ,
\end{equation}
where $Q_0(\omega)$ denotes the frequency dependent linear response of the system and $\mu$ is the chemical potential. We estimate $\frac{\partial  Q_0}{\partial N}\sim Q_0/N_0$, $\frac{\partial \Delta}{\partial \mu}\sim \Delta/E_F$,
$\frac{\partial \mu}{\partial N}\sim \lambda_{TF}^2e^2$, where $E_F$ is the Fermi energy, $\lambda_{TF}$ is the Thomas-Fermi screening length. Finally, we get
\begin{equation}
\label{kernel}
Q_{1}(\omega) \sim Q_0+ \Delta\frac{\partial Q_0}{\partial\Delta}(k_F^2\lambda_{TF}^2)\frac{k_Fe^2}{E_F}\ ,
\end{equation}
where $k_F$ is the Fermi momentum. The first term in this expression is similar to the one obtained by Hertel in \cite{Hertel1} for the normal metals though, of course,  it takes account of the full response of both nonsuperconducting and superconducting carriers. The second term is specific for superconductors and reflects the concentration dependence of the superconducting gap function. Clearly this term is nonzero only provided we take account of the gap dynamics directly induced by the incident electromagnetic wave.
Despite the presence of the small parameter $\Delta/E_F$ in one of the two contributions to the Eq.~(\ref{kernel}) the final estimate shows that both contributions to the $Q_1$ value can be comparable though their relation depends certainly on temperature.
Specifically, at temperature slightly below the superconducting transition temperature $T_c$ the dependence $Q_0(\Delta)$ is power-law and $\Delta\left(\partial Q_0/\partial \Delta\right)\sim Q_s$ where $Q_s$ is the part of the response function determined by the contribution of superconducting carriers. At low temperatures $T\rightarrow 0$ the derivative $\partial Q_0/\partial \Delta$ vanishes and, thus, the maximum value of the second term in Eq.~(\ref{kernel}) is reached at intermediate temperatures. Taking now the expression for
$\delta N(\omega)=-{\rm div} {\bf j}/i\omega e=-{\rm div} \left[Q_0c{\bf E}(\omega)/\omega^2 e\right]$
from the continuity equation one can get the nonlinear contribution to the current density at zero frequency
${\bf j}_1={\rm rot} {\bf M}_0$, where the magnetic moment is given by the vector product
\begin{equation}\label{M0}
{\bf M}_0 = -\frac{i[Q_0{\bf E}\times Q_1^*{\bf E}^*]c^2}{N_0e^2\omega^3}.
\end{equation}
It is important to note that here the electric field ${\bf E}$ is taken inside the sample and its relation to the external field of the incident electromagnetic wave should be found from the solution of the linear scattering problem.

The above simple reasoning shows that the qualitatively new physics of the IFE in superconductors can arise only from the gap modulation effect associated with the second term in Eq.~(\ref{kernel}). In order to separate this effect from the other possible 
contributions we choose a specific sample geometry, namely a thin disc of the radius $R$ much smaller than both the superconducting screening length and the light wavelength (see Fig.~\ref{Fig1}). For simplicity we also assume this radius to be less than the length $l_E$ of the relaxation of the electron-hole imbalance potential, neglecting, thus, the possible conversion between the superconducting and normal currents. 
For the quantitative consideration of the gap modulation effect we take the simplest phenomenological model known to describe the dynamics of the superconducting order parameter $\psi$ at rather low frequencies and based on the so called
 time-dependent Ginzburg-Landau (GL) equation
\begin{equation}\label{GL_general}
\left(\pi \alpha/8+i\gamma\right)\hbar\partial_t \psi+\alpha T_c\epsilon\psi+\xi_0^2{\bf \hat{D}}^2\psi+b\left|\psi\right|^2\psi=0,
\end{equation}
where ${\bf \hat{D}}=-i\nabla+(2\pi/\Phi_0){\bf A}$, $\Phi_0=\pi\hbar c/e$ ($e>0$) is the magnetic flux quantum, $\xi_0$ is the superconducting zero-temperature coherence length, and $\epsilon=T/T_c-1$. Considering the electric field ${\bf E}=E_0{\rm Re}\left[({\bf e}_x+i{\bf e}_y)e^{-i\omega t}\right]$ inside the disk we may choose the corresponding vector potential in the form ${\bf A}=(cE_0/\omega){\rm Re}\left[({\bf e}_y-i{\bf e}_x)e^{-i\omega t}\right]$ (here we choose the origin of the coordinate system with the in-plane axes $x$ and $y$ in the disk center). Here we assume the disk thickness $L$ to be much smaller than the skin depth and neglect the variation of the vector potential along the $z$ axis.
Though the above model has a rather restricted range of validity  and assumes a gapless superconducting state
 such consideration is known to provide instructive insights for a great variety of dynamic phenomena
  (see \cite{Kopnin, Larkin} for review).
 The key ingredient of Eq.~(\ref{GL_general}) responsible for the IFE  is the imaginary part of the dimensionless relaxation constant $\gamma\sim\alpha(T_c/E_F)$ arising due to electron-hole asymmetry \cite{Larkin}. In addition, one needs to impose the boundary conditions at the disk edge which guarantee $\partial \left|\psi\right|/\partial r=0$ in the cylindrical coordinates ($r$, $\theta$) and the absence of the radial superconducting current at $r=R$. The solution of Eq.~(\ref{GL_general}) determines the superconducting current flowing in the disk:
\begin{equation}\label{js_def}
{\bf j}_s=\frac{2\pi\alpha T_c\xi_0^2 c}{\Phi_0}~\left(i\psi^*\nabla\psi -i\psi\nabla \psi^*-\frac{4\pi}{\Phi_0}\left|\psi\right|^2{\bf A}\right).
\end{equation}
Note that since $R\ll l_E$ Eq.~(\ref{GL_general}) does not contain the electrochemical potential.

\begin{figure}[t!]
\includegraphics[width=0.48\textwidth]
{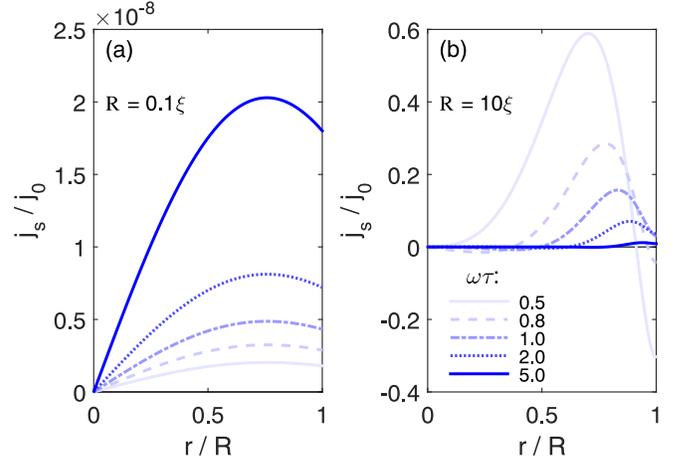}
\caption{Dependencies of the azimuthal superconducting current as a function of the distance $r$ from the disk center for (a) $R=0.1\xi$ and (b) $R=10\xi$. Different curves correspond to different radiation frequencies. The values of the parameter $\omega\tau$ relevant for both panels are shown in the panel (b). We denote $j_0=32\gamma  e^3\Delta_0^2E_0^2\tau^3/\left(\pi^2\alpha^2 m^2 \xi\right)$.} \label{Fig2}
\end{figure}

\begin{figure*}[t!]
\includegraphics[width=0.48\textwidth]
{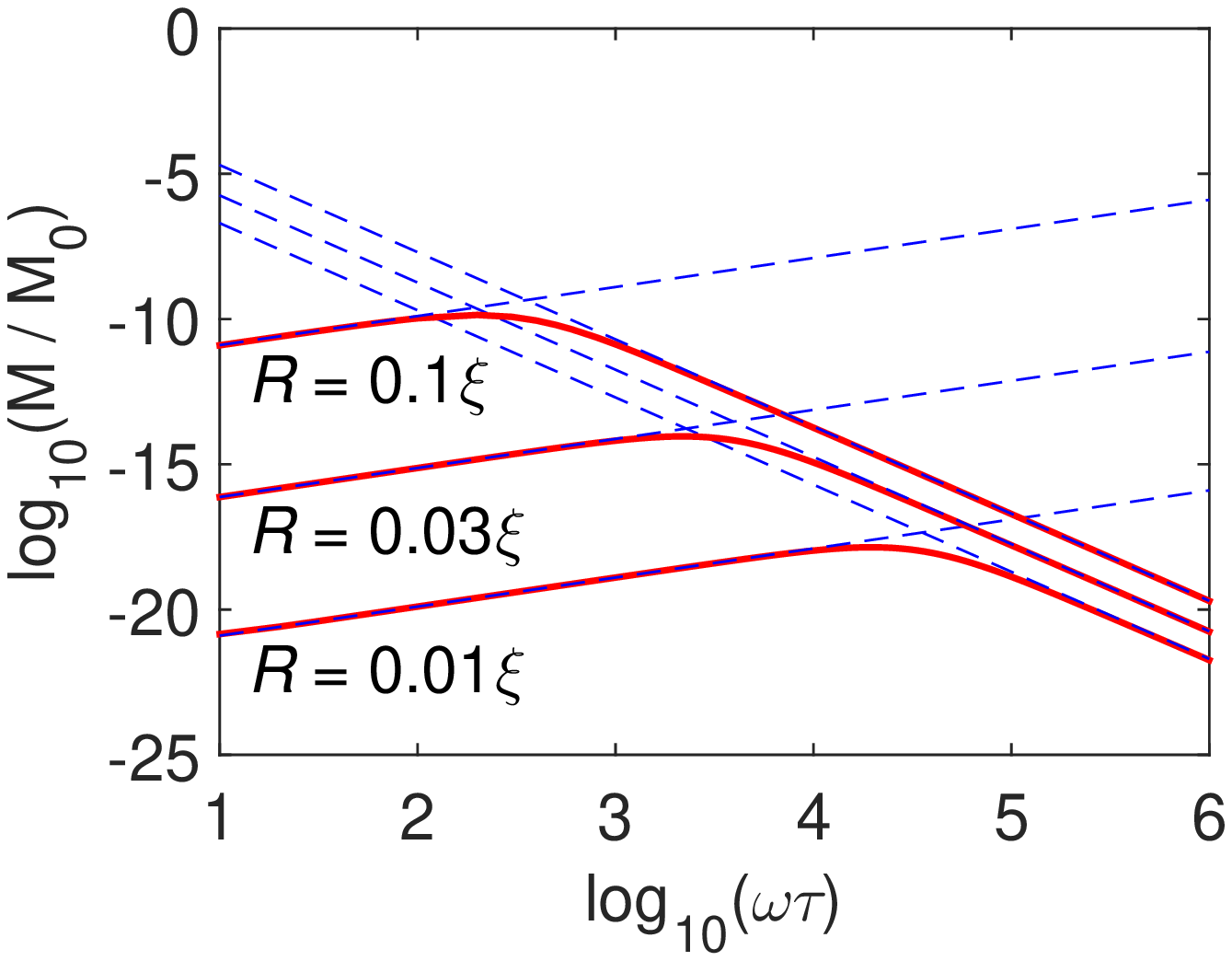}
\includegraphics[width=0.48\textwidth]{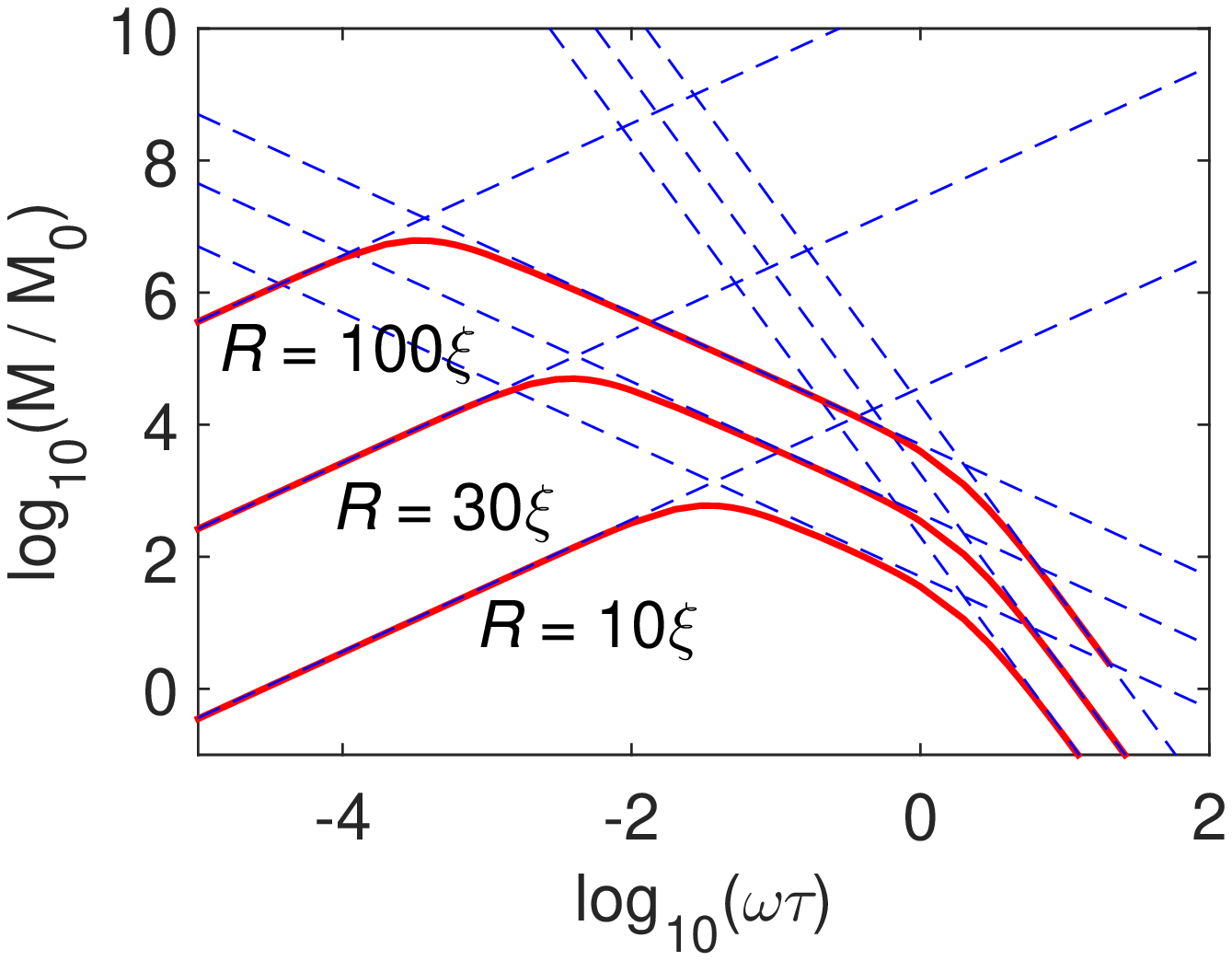}
\caption{Dependence of the magnetic moment on the radiation frequency for the disks with (a) $R\ll \xi$ and (b) $R\gg\xi$. Here $M_0=8\pi^4\gamma \hbar \tau\xi^6 c^2  L\Delta_0^2E_0^2/\Phi_0^3
$.} \label{Fig_MOmega}
\end{figure*}

Further it is  convenient to introduce the absolute value $\Delta$ and the phase $\chi$ of the order parameter: $\psi=\Delta\exp(i\chi)$.
In the absence of radiation the superconducting gap can be chosen real and equal to $\Delta_0^2=-(\alpha T_c/b)\epsilon$. The vector potential of the incident wave results in the corrections $\Delta_1\propto A$ and $\chi\propto A$ to the gap value and the superconducting phase, respectively. Introducing the temperature-dependent correlation length $\xi=\xi_0/\sqrt{|\epsilon|}$ ($\epsilon<0$),  the GL time $\tau=\pi\hbar/8T_c|\epsilon|$ and the small parameter $\nu=8\gamma/\pi\alpha\propto(T_c/E_F)\ll 1$ we obtain the first-order equations for $\Delta_1$ and $\chi$ together with the boundary conditions:
\begin{equation}\label{GL_re_lin}
\tau\frac{\partial \Delta_1}{\partial t}-\nu\tau\Delta_0\frac{\partial \chi}{\partial t}+2\Delta_1-\xi^2\nabla^2\Delta_1=0,
\end{equation}
\begin{equation}\label{GL_im_lin}
\tau\Delta_0\frac{\partial \chi}{\partial t}+\nu\tau\frac{\partial \Delta_1}{\partial t}-\xi^2\Delta_0\nabla^2\chi=0.
\end{equation}
\begin{equation}\label{BC}
\left.\frac{\partial\Delta_1}{\partial r}\right|_{r=R}=0,~~~\left.\left(\frac{\partial\chi}{\partial r}+\frac{2\pi}{\Phi_0}A_r\right)\right|_{r=R}=0.
\end{equation}
Note that the vector potential ${\bf A}$ controls only the boundary condition. Since the system (\ref{GL_re_lin})-(\ref{BC}) is linear it is convenient to introduce the complex amplitudes $\tilde u$: $u={\rm Re}\left(\tilde{u}e^{i\theta-i\omega t}\right)$, where $u=\left\{\Delta_1,~\chi,~{\bf A}\right\}$. Then the solution for the $\tilde{\Delta_1}$ and $\tilde{\chi}$ which accounts only the corrections up to $O(\nu)$ reads \cite{supp}:
\begin{equation}\label{BC_chi}
\tilde{\chi}(r)=\frac{2\pi i R}{\Phi_0}\frac{cE_0}{\omega}f(q_2,r),
\end{equation}
\begin{equation}\label{Delta_res}
\tilde{\Delta}_1(r)=\Delta_0\frac{\pi RcE_0\nu\tau}{\Phi_0}\left[f(q_2,r)-f(q_1,r)\right].
\end{equation}
Here we have introduced the values $q_1=\xi^{-1}\left(\sqrt{\sqrt{1+w^2}-1}+i\sqrt{\sqrt{1+w^2}+1}\right)$, $q_2=\xi^{-1}\sqrt{2iw}$, $w=\omega\tau/2$ and
\begin{equation}\label{f_def}
f(q,r)=\frac{J_1(qr)}{qRJ_0(qR)-J_1(qR)},
\end{equation}
where $J_0$ and $J_1$ are the Bessel functions.

The obtained complex amplitudes enable us to calculate the spatial profile of the d.c.  superconducting current:
\begin{equation}\label{Curr_complex}
\left<j_{s\theta}(r)\right>=\frac{4\pi \alpha T_c\xi_0^2 c}{\Phi_0}\Delta_0{\rm Re}\left\{\tilde{\Delta}_1\left(\frac{i\tilde{\chi}^*}{r}-\frac{2\pi}{\Phi_0}\tilde{A}_\theta^*\right)\right\}.
\end{equation}
After substitution and simplification we obtain:
\begin{equation}\label{Curr_res}
\begin{array}{c}{\ds
\left<j_{s\theta}(r)\right>=\frac{\gamma}{\alpha}\frac{32  e^3\Delta_0^2E_0^2\tau^2}{\pi^2\alpha m^2 \omega R}\left(\frac{R}{\xi}\right)^2\times}\\{}\\{\ds \times{\rm Re}\left\{\left[f(q_1,r)-f(q_2,r)\right]\left[1-(R/r)f(q_2^*,r)\right]\right\},}
\end{array}
\end{equation}
where $m=\hbar^2/(4\alpha T_c\xi_0^2)$ is the mass characterizing the Cooper pair.

The typical profiles of the d.c. current in the disk are shown in Fig.~\ref{Fig2}. Remarkably, for relatively large disks the current changes its direction at the certain distances from the center. Qualitatively, this phenomenon is associated with the coupled oscillations of the amplitude and phase of the superconducting order parameter [see Eqs. (\ref{GL_re_lin})-(\ref{BC})]. Note that in the limit $\gamma\gg\alpha$ these oscillations take the form of the sound-like waves similar to the Bogoliubov modes in the Bose-Einstein condensate (see, e.g., \cite{Pethick}). The oscillations of the order parameter amplitude are responsible for the appropriate contribution to the supercurrent and it is the sign change of the correction $\Delta_1$ which determines the sign change to the supercurrent (\ref{Curr_complex}).  At the same time, the total magnetic moment of the disk $M=(L/c)\int_0^R\left<j_{s\theta}\right>\pi r^2dr$ (where $L$ is the disk thickness) is fixed by the light polarization (see \cite{supp} for the analytical expression). The typical dependencies of the total magnetic moment 𝑀 on the radiation frequency $\omega$ for
different ratios $(R/\xi)$ are shown in Fig.~\ref{Fig_MOmega}.

For small disks with $R\ll \xi$ the magnetic moment $M$ linearly grows as a function of $\omega$ in the low-frequency limit and reveals a crossover to $M\propto \omega^{-3}$ behavior at $\omega\tau\sim(\xi/R)^2$:
\begin{equation}\label{M_small_R}
M=\frac{\gamma}{\alpha}\frac{64 e^3\Delta_0^2E_0^2LR^2}{\pi\alpha m^2 c}\left\{\begin{array}{l}{\ds \beta\omega\tau^4\left(\frac{R}{\xi}\right)^8,~{\rm for}~\omega\tau\ll\left(\frac{\xi}{R}\right)^2,}\\{}\\{\ds \frac{1}{\omega^3},~~~~~~~~~~~~~{\rm for}~\omega\tau\gg\left(\frac{\xi}{R}\right)^2,}
\end{array}\right.
\end{equation}
where $\beta=73/11520\approx 6\cdot 10^{-3}$.

In the opposite limit when $R\gg\xi$ the dependence $M(\omega)$ is characterized by three regimes:
\begin{equation}\label{M_large_R}
M=\frac{\gamma}{\alpha}\frac{64 e^3\Delta_0^2E_0^2LR^2}{\pi\alpha m^2 c}\left\{\begin{array}{l}{\ds \eta\omega\tau^4\left(\frac{R}{\xi}\right)^4,~{\rm for}~\omega\tau\ll\left(\frac{\xi}{R}\right)^2,}\\{}\\{\ds \frac{\tau^2}{4\omega},~~~~~{\rm for}~\left(\frac{\xi}{R}\right)^2\ll \omega\tau\ll 1,}\\{}\\{\ds \frac{1}{\omega^3},~~~~~{\rm for}~\omega\tau\gg 1,}
\end{array}\right.
\end{equation}
where $\eta=7/384\approx 0.018$.

Note that Eq.~(\ref{GL_general}) is valid for the radiation frequency $\omega$ smaller than the characteristic frequencies of the inelastic electron-phonon relaxation processes $\tau_{ph}^{-1}$ \cite{Kopnin}. At the same time, the GL relaxation time $\tau$ becomes infinitely large when $T$ approaches $T_c$. Thus, even in the limit $\omega\ll \tau_{ph}^{-1}$ the parameter $\omega\tau$ in (\ref{M_small_R})-(\ref{M_large_R}) can take the values both greater and smaller than unity.

The above contribution to the magnetic moment $M$ of the disk should be, of course, summed up with the term arising from the Hertel contribution associated with the non-superconducting electrons \cite{Hertel1}.

It is important to note that the electric field ${\bf E}$ acting on the electrons {\it inside} the superconducting disk does not coincide with the electric field ${\bf E}^{ext}$ of the incident electromagnetic wave due to the depolarization effects. The simplified relation between these quantities can be established by approximating the disk with the conducting ellipsoid of the semi-axes $R$, $R$, and $L/2$. In this case, one gets \cite{Landau}
\begin{equation}
{\bf E}=\frac{{\bf E}^{ext}}{1+(\pi/8)(L/R)\left[\varepsilon(\omega)-1\right]},
\end{equation}
where $\varepsilon(\omega)$ is the frequency dependent permittivity of the superconducting metal. In the collisionless limit $\varepsilon(\omega)=1-\omega_p^2/\omega^2$ where $\omega_p=(4\pi e^2 N_0/m)^{1/2}$ is the electronic plasma frequency of the metal. As a result, in the low-frequency limit when $\omega\ll\omega_p\sqrt{L/R}$ in Eq.~(\ref{M0}) one finds $\left|{\bf E}\right|^2\sim (\omega/\omega_p)^4(R/L)^2\left|{\bf E}^{ext}\right|^2$. This additional factor coming from the depolarization effects provides a natural cut-off for the the expression (\ref{M0}) in the limit $\omega\to 0$. 
Because of this renormalization the magnetization arising in normal metals due to IFE (see \cite{Hertel1}) should also vanish at zero frequency.

To estimate the possible values of the optically induced magnetic moment for simplicity let us consider the case $R\gtrsim \xi$ and $\omega\tau\gg 1$. Then accounting that $\gamma/\alpha\sim T_c/E_F$ \cite{Larkin}, $\alpha\sim T_c/E_F$ from the expression (\ref{M_large_R}) we obtain
\begin{equation}
M\sim 10\mu_B\frac{\Delta_0^2}{N}\frac{P}{\hbar \omega^2}.
\end{equation}
Here $\mu_B=e\hbar/(2mc)$ is the Bohr magneton, $P=(\omega_p^2/\omega)(\left| {\bf E}\right|^2V/4\pi)$ is the total power absorbed inside the superconducting disk, $V=\pi R^2L$ is the disk volume, and $\omega_p=(4\pi e^2 N/m)^{1/2}$ is the electron plasma frequency. The relation between $P$ and the electromagnetic wave intensity ${\mathcal I}=c\left| {\bf E}^{ext}\right|^2/8\pi$ is determined by the sample geometry and the radiation frequency. For $\omega^2\ll \omega_p^2(L/R)$ the depolarization effects significantly influences the electric field inside the disk so that $\left| {\bf E}\right|\sim \left| {\bf E}^{ext}\right| (\omega^2/\omega_p^2)(R/L)$. As a result, at temperatures $T\ll T_c$ we may put $\Delta_0^2\approx N$ and finally obtain
\begin{equation}
M\sim 10\mu_B\frac{\omega^2}{\omega_p^2}\frac{R^2}{L^2}\frac{{\mathcal I}V}{\hbar \omega c}.
\end{equation}
Taking the disk of the radius $R\sim 10~{\rm \mu m}$ and the thickness $L\sim 10~{\rm nm}$ radiated by the infra-red wave with $\omega \sim 10^{13}~{\rm sec^{-1}}$ and intensity ${\mathcal I}\sim 10~ {\rm  \mu W/\mu m^2}$, and taking $\omega_p\sim 10^{15}~{\rm sec^{-1}}$ (keeping in mind high-$T_c$ cuprates) we can get $M\sim 10^3 \mu_B$.

Of course, choosing the appropriate laser beam intensity one needs to make sure that the heating does not destroy the superconducting state. To arrange the effective heat removal one can, e.g., place the superconducting film on top of the sapphire substrate. Thanks to the very high thermal conductivity of sapphire $\kappa\sim 10^3~{\rm W/(m\cdot K)}$ at $T\sim 10~K$ \cite{Berman}, the substrate of the thickness $d_{sub}\sim 1~{\rm \mu m}$ with the temperature elevation of $\Delta T\sim 1~{\rm K}$ between its edges can support the heat transfer of the surface power density up to $q\sim \kappa \Delta T/d_{sub}\sim10^3~{\rm \mu W}/{\rm \mu m}^2$.  Thus, even in the case of the full absorption of the laser radiation of the intensity ${\mathcal I}\sim 10~ {\rm  \mu W/\mu m^2}$ the disk on top of the substrate remains superconducting. The experimental achievability of such regime was demonstrated, e.g., in Ref.~\cite{Veshchunov} where under the influence of the laser beam with the absorbed power $17~{\rm \mu W}$ the temperature of Nb film of the thickness $90~{\rm nm}$ placed on the $500~{\rm \mu m}$ Si substrate did not exceed $T_c$.

Clearly, the predicted phenomena should become dramatically enhanced in superconducting media where $T_c/E_F\sim 1$ (e.g., Bose-Einstein condensates in the local pairs condensation regime \cite{Ketterie} or the superconductors with extremely small electron density like SrTiO$_3$ \cite{Behnia}). In such systems the superconducting currents induced by the circularly polarized light may become large enough to generate vortices which can survive due to the presence of the pinning effects even after the laser pulse is switched off.

Finally, we note that the measurements of the IFE contribution caused by the gap modulation provide an interesting possibility to establish the connection between three different phenomena: IFE, the direct Faraday effect, and the Hall effect in the vortex state. Indeed, the imaginary part of the relaxation constant $\gamma$ is known to affect the vortex contribution to the off-diagonal component $\sigma_{xy}$ of the complex conductivity tensor which controls the rotation of the light polarization plane due to the direct Faraday effect \cite{Dorsey, Kopnin_Hall}. The renormalization of $\sigma_{xy}$ due to the non-zero $\gamma$ can be also responsible for the anomalous sign change of the Hall coefficient near the superconducting critical temperature for a number of superconducting compounds including high-$T_c$ cuprates. The sign of the $\gamma$ parameter determines, thus, both the sign of the contributions to the induced magnetic moment in the IFE and to the Hall conductivity governed by the gap dynamics.

\acknowledgements

This work was supported by the Russian Science Foundation (Grant No. 18-72-10027) in  part related to the calculation of magnetic moment arising due to IFE and the Russian Foundation for Basic Research
(Grant No. 18-02-00390). S.M. acknowledges the financial support of the  Foundation for the Advancement of Theoretical Physics and Mathematics “BASIS” (grant No. 18-1-3-58-1) and Russian Presidential Scholarship SP-3938.2018.5. A.S.M. acknowledges support from the Bordeaux University IDEX program. The work of A.B. and B.L. was supported by the French ANR OPTOFLUXONICS, EU COST CA16218 Nanocohybri.

\renewcommand{\theequation}{S\arabic{equation}}

\begin{widetext}

\section*{Supplemental material for ``Inverse Faraday Effect for Superconducting Condensates'': Solution of time-dependent Ginzburg-Landau equation}

\setcounter{equation}{0}

To find the solution of Eqs.~(6)-(7) with the boundary conditions (8) from the main text of the paper it is convenient to introduce the complex amplitudes $\tilde u$: $u={\rm Re}\left(\tilde{u}e^{i\theta-i\omega t}\right)$, where $u=\left\{\Delta_1,~\chi,~{\bf A}\right\}$. We will be interested only in the first-order corrections with respect to the electric field amplitude. Since all these corrections are proportional to $\exp(i\theta)$ we need to find only their dependencies on $r$. To do this let us search the solution of the Ginzburg-Landau equations proportional to the first order Bessel function: $\Delta_1=\Delta_0 \delta_q J_1(qr)$, $\chi=\chi_q J_1(qr)$. Then the equations become algebraic:
\begin{equation}\label{GL_re_lin_Bessel}
i\nu\omega\tau\chi_q+(2+q^2\xi^2-i\omega\tau)\delta_q=0,
\end{equation} 
\begin{equation}\label{GL_im_lin_Bessel}
(i\omega\tau-q^2\xi^2)\chi_q+i\nu\omega\tau\delta_q=0.
\end{equation} 
This system has non-trivial solutions for $\chi_q$ and $\delta_q$ only if the parameter $q$ satisfies the equation
\begin{equation}\label{q_eq}
(2+q^2\xi^2-i\omega\tau)(i\omega\tau-q^2\xi^2)+\nu^2\omega^2\tau^2=0.
\end{equation} 
In what follows we neglect the corrections $\propto \nu^2$ to the $q$ values. Then Eq.~(\ref{q_eq}) splits into two separate quadratic equations. Since the function $J_1(z)$ is odd in the complex plane we may consider only one root of each quadratic equation (other ones do not produce new physical solutions for $\chi_q$ and $\delta_q$):
\begin{equation}\label{q_solve}
q_1=\xi^{-1}\left(\sqrt{\sqrt{1+\left(\frac{\omega\tau}{2}\right)^2}-1}+i\sqrt{\sqrt{1+\left(\frac{\omega\tau}{2}\right)^2}+1}\right), ~~~~~q_2=(1+i)\xi^{-1}\sqrt{\frac{\omega\tau}{2}}.
\end{equation} 

The phase of the order parameter is defined by the boundary condition (8) from the main text of the paper. Neglecting the corrections $\propto \nu$ we obtain:
\begin{equation}\label{BC_chi}
\chi(r)=\frac{2\pi i R}{\Phi_0}\frac{cE_0}{\omega}\frac{J_1(q_2r)}{q_2RJ_0(q_2R)-J_1(q_2R)}.
\end{equation}
Then the general solution of the equation for $\Delta_1$ reads
\begin{equation}\label{BC_Delta}
\Delta_1(r)=\Delta_0\left[\delta_qJ_1(q_1r)-\frac{i\nu\omega\tau}{2+q_2^2\xi^2-i\omega\tau}\frac{2\pi i }{\Phi_0}\frac{cE_0}{\omega}\frac{RJ_1(q_2r)}{q_2RJ_0(q_2R)-J_1(q_2R)}\right].
\end{equation}
Then using the boundary condition for $\Delta_1$ and finding $\delta_q$ we get:
\begin{equation}\label{BC_Delta_2}
\delta_q=\frac{i\nu\omega\tau}{2}\frac{2\pi i }{\Phi_0}\frac{cE_0}{\omega}\frac{R}{q_1RJ_0(q_1R)-J_1(q_1R)}
\end{equation}
\begin{equation}\label{Delta_res}
\Delta_1(r)=-\Delta_0\frac{\nu\omega\tau}{2}\frac{2\pi R}{\Phi_0}\frac{cE_0}{\omega}\left[\frac{J_1(q_1r)}{q_1RJ_0(q_1R)-J_1(q_1R)}-\frac{J_1(q_2r)}{q_2RJ_0(q_2R)-J_1(q_2R)}\right].
\end{equation}

The obtained corrections enable us to calculate the time-averaged superconducting current:
\begin{equation}\label{Curr_corr}
\left<j_{s\theta}\right>=-\frac{8\pi \alpha T_c\xi_0^2 c}{\Phi_0}\Delta_0\left<\Delta_1\left(\frac{1}{r}\frac{\partial\chi}{\partial\theta}+\frac{2\pi}{\Phi_0}A_\theta\right)\right>.
\end{equation}
In terms of complex amplitudes this expression takes the form
\begin{equation}\label{Curr_complex}
\left<j_{s\theta}\right>=-\frac{4\pi \alpha T_c\xi_0^2 c}{\Phi_0}\Delta_0{\rm Re}\left\{\Delta_1\left(-\frac{i\chi^*}{r}+\frac{2\pi}{\Phi_0}A_\theta^*\right)\right\}.
\end{equation}
After substitution and simplification we obtain:
\begin{equation}\label{Curr_res}
\begin{array}{c}{\ds 
\left<j_{s\theta}\right>=\frac{\gamma}{\alpha}\frac{8\pi^3\hbar\alpha \xi^2 c^3 R\Delta_0^2E_0^2}{\Phi_0^3\omega}{\rm Re}\left\{\left[\frac{J_1(q_1r)}{q_1RJ_0(q_1R)-J_1(q_1R)}-\frac{J_1(q_2r)}{q_2RJ_0(q_2R)-J_1(q_2R)}\right]\times\right.}\\{}\\{\ds \left.\times\left[1-\frac{R}{r}\frac{J_1(q_2^*r)}{q_2^*RJ_0(q_2^*R)-J_1(q_2^*R)}\right]\right\}.}
\end{array}
\end{equation}
The magnetic moment of the disk reads
\begin{equation}\label{m_def}
M=\frac{L}{c}\int_0^R\left<j_{s\theta}\right>\pi r^2dr,
\end{equation}
where $L$ is the thickness of the superconducting disk. To simplify the form of the answer let us introduce the values $x_1=q_1R$, $x_2=q_2R$ and $x_2^*=q_2^*R$. Then after calculations we obtain:
\begin{equation}\label{m_res}
M=\frac{\gamma}{\alpha}\frac{8\pi^4\hbar\alpha \xi^2 c^2 R^4L\Delta_0^2E_0^2}{\Phi_0^3\omega}Z,
\end{equation}
where for convenience we introduce the value 
\begin{equation}\label{Q_def}
Z=Z_1+Z_2
\end{equation}
with
\begin{equation}\label{Q1_def}
Z_1={\rm Re}\left[\frac{1}{x_1}\frac{J_2(x_1)}{x_1J_0(x_1)-J_1(x_1)}-\frac{1}{x_2}\frac{J_2(x_2)}{x_2J_0(x_2)-J_1(x_2)}\right]
\end{equation}
and
\begin{equation}\label{Q2_def}
Z_2={\rm Re}\left\{\frac{1}{x_2^*J_0(x_2^*)-J_1(x_2^*)}\left[\frac{x_2^{*}J_0(x_2^*)J_1(x_2)-x_2J_0(x_2)J_1(x_2^*)}{(x_2^2-x_2^{*2})\left[x_2J_0(x_2)-J_1(x_2)\right]}-\frac{x_2^*J_0(x_2^*)J_1(x_1)-x_1J_0(x_1)J_1(x_2^*)}{(x_1^2-x_2^{*2})\left[x_1J_0(x_1)-J_1(x_1)\right]}\right]\right\}.
\end{equation}

\subsection{The limit $R/\xi\ll 1$ and $\omega\tau\ll (\xi/R)^2$ }

In this limit the absolute values of all arguments of the Bessel functions are small. To obtain the non-trivial result we need to expand the above expression for the magnetic moment up to $O[(R/\xi)^6]$.  Note that one has the {\it exact} expressions
\begin{equation}\label{x_simp}
x_2^2=i\omega\tau\left(\frac{R}{\xi}\right)^2,~~~x_1^2= \left(-2+i\omega\tau\right)\left(\frac{R}{\xi}\right)^2
\end{equation}
Let us consider different parts of the answer. For $x\ll 1$ 
\begin{equation}\label{s1}
\frac{1}{x}\frac{J_2(x)}{xJ_0(x)-J_1(x)}=\frac{1}{4}+\frac{7}{96}x^2+\frac{11}{512}x^4+\frac{73}{11520}x^6+O(x^8)
\end{equation} 
Then
\begin{equation}\label{s2}
Z_1= {\rm Re}\left[\frac{7}{96}(x_1^2-x_2^2)+\frac{11}{512}(x_1^4-x_2^4)+\frac{73}{11520}(x_1^6-x_2^6)\right]
\end{equation}
Now we turn to the second part of the answer. Expanding it up to the terms $\sim O[(R/\xi)^6]$ we find:
\begin{equation}\label{s3}
\begin{array}{c}{\ds
\frac{x_2^{*}J_0(x_2^*)J_1(x)-xJ_0(x)J_1(x_2^*)}{(x^2-x_2^{*2})\left[x_2^*J_0(x_2^*)-J_1(x_2^*)\right]\left[xJ_0(x)-J_1(x)\right]}=}\\{}\\{\ds =\left(\frac{1}{4}+\frac{7}{96}x_2^{*2}+\frac{11}{512}x_2^{*4}+\frac{73}{11520}x_2^{*6}\right)+\left(\frac{7}{96}+\frac{11}{512}x_2^{*2}+\frac{73}{11520}x_2^{*4}\right)x^2+\left(\frac{11}{512}+\frac{73}{11520}x_2^{*2}\right)x^4+\frac{73}{11520}x^6}
\end{array}
\end{equation}
Then we find:
\begin{equation}\label{s4}
Z_2 ={\rm Re}\left[\left(\frac{7}{96}+\frac{11}{512}x_2^{*2}+\frac{73}{11520}x_2^{*4}\right)(x_2^2-x_1^2)+\left(\frac{11}{512}+\frac{73}{11520}x_2^{*2}\right)(x_2^4-x_1^4)+\frac{73}{11520}(x_2^6-x_1^6)\right]
\end{equation}
This gives us the expression for $Z=Z_1+Z_2$:
\begin{equation}\label{s5}
Z={\rm Re}\left[\left(\frac{11}{512}x_2^{*2}+\frac{73}{11520}x_2^{*4}\right)(x_2^2-x_1^2)+\frac{73}{11520}x_2^{*2}(x_2^4-x_1^4)\right]
\end{equation}
Substituting the expressions for $x_1$ and $x_2$ into Eq.~(\ref{s5}) after simplifications we obtain:
\begin{equation}\label{s6}
Z=\frac{73}{5760}\omega^2\tau^2\left(\frac{R}{\xi}\right)^6.
\end{equation}

\subsection{The limit $R/\xi\ll 1$ and $\omega\tau\gg (\xi/R)^2\gg 1$ }

In this limit the expansion of the value $x_1$ reads
\begin{equation}\label{x1_large}
x_1=x_2-\frac{1-i}{\sqrt{2\omega\tau}}\frac{R}{\xi}+\frac{1+i}{(2\omega\tau)^{3/2}}\frac{R}{\xi}+O\left\{\left(\frac{1}{\omega\tau}\right)^{5/2}\right\}
\end{equation}
Let us expand the resulting expression for the magnetic moment over $\Delta x=x_1-x_2$. To do this it is convenient to introduce the function
\begin{equation}\label{F_def}
F(x_1)=\frac{1}{x_1}\frac{J_2(x_1)}{x_1J_0(x_1)-J_1(x_1)}-\frac{1}{x_2^*J_0(x_2^*)-J_1(x_2^*)}\frac{x_2^*J_0(x_2^*)J_1(x_1)-x_1J_0(x_1)J_1(x_2^*)}{(x_1^2-x_2^{*2})\left[x_1J_0(x_1)-J_1(x_1)\right]}
\end{equation}
Then
\begin{equation}\label{Q_def}
Z\approx {\rm Re}\left\{\left.\frac{\partial F(x_1)}{\partial x_1}\right|_{x_1=x_2}\Delta x+\frac{1}{2}\left.\frac{\partial^2 F(x_1)}{\partial x_1^2}\right|_{x_1=x_2}\Delta x^2\right\}
\end{equation}
After simplifications we get:
\begin{equation}\label{F_2}
F(x_1)=\frac{2x_1^2J_2(x_2^*)\left[J_0(x_1)-J_2(x_1)\right]-2x_2^{*2}J_2(x_1)\left[J_0(x_2^*)-J_2(x_2^*)\right]}{x_1^2(x_1^2-x_2^{*2})\left[J_0(x_1)-J_2(x_1)\right]\left[J_0(x_2^*)-J_2(x_2^*)\right]}
\end{equation}

Let us analyze the properties of the function $F(x_1)$ in more detail. For $\left|x\right|\gg 1$ one gets

\begin{equation}
J_2(x)=-\frac{15}{4\sqrt{2}\sqrt{\pi}}\frac{1}{x^{3/2}}\cos\left(\frac{\pi}{4}+x\right)-\left[\frac{\sqrt{2}}{\sqrt{\pi}}\frac{1}{x^{1/2}}-\frac{105}{64\sqrt{2\pi}}\frac{1}{x^{5/2}}\right]\sin\left(\frac{\pi}{4}+x\right)+O\left(\frac{1}{x^{7/2}}\right)
\end{equation}
\begin{equation}
J_0(x)-J_2(x)=\frac{7}{2\sqrt{2\pi}}\frac{1}{x^{3/2}}\cos\left(\frac{\pi}{4}+x\right)+\left[\frac{2\sqrt{2}}{\sqrt{\pi}}\frac{1}{x^{1/2}}-\frac{57}{32\sqrt{2\pi}}\frac{1}{x^{5/2}}\right]\sin\left(\frac{\pi}{4}+x\right)+O\left(\frac{1}{x^{7/2}}\right)
\end{equation}
To proceed we extract explicitly the real and imaginary parts: $x_1=u_1+iv_1$, $x_2=u_2+iv_2$, and $x_2^*=u_2-iv_2$. It is important that  $v_1>0$ and $v_2>0$. Then assuming that $x$ takes one of the values $x_1$ or $x_2$ with the exponential accuracy we find:
\begin{equation}
J_2(x)=-\frac{i}{\sqrt{2\pi}x^{1/2}}\left(1-\frac{15}{8}\frac{i}{x}-\frac{105}{128}\frac{1}{x^2}\right)e^ve^{-i\left(\pi/4+u\right)}
\end{equation}
\begin{equation}
J_0(x)-J_2(x)=i\frac{\sqrt{2}}{\sqrt{\pi}}\frac{1}{x^{1/2}}\left(1-\frac{7}{8}\frac{i}{x}-\frac{57}{128}\frac{1}{x^{2}}\right)e^ve^{-i\left(\pi/4+u\right)}
\end{equation}
\begin{equation}
J_2(x_2^*)=\frac{i}{\sqrt{2\pi}x_2^{*1/2}}\left(1+\frac{15}{8}\frac{i}{x_2^*}-\frac{105}{128}\frac{1}{x_2^{*2}}\right)e^{v_2}e^{i\left(\pi/4+u_2\right)}
\end{equation}
\begin{equation}
J_0(x_2^*)-J_2(x_2^*)=-i\frac{\sqrt{2}}{\sqrt{\pi}}\frac{1}{x_2^{*1/2}}\left(1+\frac{7}{8}\frac{i}{x_2^*}-\frac{57}{128}\frac{1}{x_2^{*2}}\right)e^{v_2}e^{i\left(\pi/4+u_2\right)}
\end{equation}
Then
\begin{equation}\label{F_4}
F(x)=-\frac{x^2\left(1+\frac{15}{8}\frac{i}{x_2^*}-\frac{105}{128}\frac{1}{x_2^{*2}}\right)\left(1-\frac{7}{8}\frac{i}{x}-\frac{57}{128}\frac{1}{x^{2}}\right)-x_2^{*2}\left(1-\frac{15}{8}\frac{i}{x}-\frac{105}{128}\frac{1}{x^2}\right)\left(1+\frac{7}{8}\frac{i}{x_2^*}-\frac{57}{128}\frac{1}{x_2^{*2}}\right)}{x^2(x^2-x_2^{*2})\left(1-\frac{7}{8}\frac{i}{x}-\frac{57}{128}\frac{1}{x^{2}}\right)\left(1+\frac{7}{8}\frac{i}{x_2^*}-\frac{57}{128}\frac{1}{x_2^{*2}}\right)}
\end{equation}

Now we are ready to calculate the full magnetic moment determined by the value $Z={\rm Re}\left[F(x_1)-F(x_2)\right]$.
\begin{equation}\label{Q_2}
\begin{array}{c}{\ds Z={\rm Re}\left\{\frac{x_2^{*2}\left(1-\frac{15}{8}\frac{i}{x_1}-\frac{105}{128}\frac{1}{x_1^2}\right)}{x_1^2(x_1^2-x_2^{*2})\left(1-\frac{7}{8}\frac{i}{x_1}-\frac{57}{128}\frac{1}{x_1^{2}}\right)}-\frac{x_2^{*2}\left(1-\frac{15}{8}\frac{i}{x_2}-\frac{105}{128}\frac{1}{x_2^2}\right)}{x_2^2(x_2^2-x_2^{*2})\left(1-\frac{7}{8}\frac{i}{x_2}-\frac{57}{128}\frac{1}{x_2^{2}}\right)}\right.}\\{}\\{\ds \left. +\left[\frac{1}{(x_2^2-x_2^{*2})}-\frac{1}{(x_1^2-x_2^{*2})}\right]\frac{\left(1+\frac{15}{8}\frac{i}{x_2^*}-\frac{105}{128}\frac{1}{x_2^{*2}}\right)}{\left(1+\frac{7}{8}\frac{i}{x_2^*}-\frac{57}{128}\frac{1}{x_2^{*2}}\right)}\right\}}
\end{array}
\end{equation}
After algebraic simplifications we get:
\begin{equation}\label{Q_3}
\begin{array}{c}{\ds Z={\rm Re}\left\{\frac{x_2^{*2}}{x_1^2(x_1^2-x_2^{*2})}\left(1-\frac{i}{x_1}+\frac{1}{2x_1^2}\right)-\frac{x_2^{*2}}{x_2^2(x_2^2-x_2^{*2})}\left(1-\frac{i}{x_2}+\frac{1}{2x_2^2}\right)\right.}\\{}\\{\ds \left. +\left[\frac{1}{(x_2^2-x_2^{*2})}-\frac{1}{(x_1^2-x_2^{*2})}\right]\left(1+\frac{i}{x_2^*}+\frac{1}{2x_2^{*2}}\right)\right\}}
\end{array}
\end{equation}
Note that $x_2^*=-ix_2$, $x_2^{*2}=-x_2^2$ and $x_1^2=x_2^2-2r^2$ where $r=R/\xi\ll 1$. Then we find:
\begin{equation}\label{Q_3}
Z={\rm Re}\left\{-\frac{1}{x_1^2}\left(1-\frac{i}{x_1}+\frac{1}{2x_1^2}\right)+\frac{1}{(x_1^2+x_2^2)}\left(-\frac{i}{x_1}+\frac{1}{2x_1^2}+\frac{1}{x_2}+\frac{1}{2x_2^2}\right)+\frac{1}{2x_2^2}\left(2-\frac{1+i}{x_2}\right) \right\}
\end{equation}
In the leading order over $1/\omega$ we finally obtain:
\begin{equation}\label{Q_4}
Z=\frac{2}{\left(\omega\tau\right)^2}\frac{\xi^2}{R^2}.
\end{equation}

\subsection{The limit $R\gg \xi$ and $\omega\tau\gg 1$}

In this limit $\omega\tau(R/\xi)^2\gg 1$ and the expression for the magnetic moment transforms in a full analogy with the corresponding limit in the case $R\ll \xi$. Thus,
\begin{equation}\label{Q_5}
Z=\frac{2}{\left(\omega\tau\right)^2}\frac{\xi^2}{R^2}
\end{equation}

\subsection{The limit $R\gg \xi$ and $\omega\tau\ll (\xi/R)^2$}

Since $R\gg\xi$ and $\omega\tau\ll (\xi/R)^2$ then $\omega\tau (R/\xi)\ll 1$. In this limit we find:
\begin{equation}
x_2=(1+i)\frac{R}{\xi}\sqrt{\frac{\omega\tau}{2}}
\end{equation}
\begin{equation}
x_1=\frac{(\omega\tau)}{2\sqrt{2}}\frac{R}{\xi}\left[1+O\left\{(\omega\tau)^2\right\}\right]+\sqrt{2}i\frac{R}{\xi}\left[1+\frac{(\omega\tau)^2}{32}+O\left\{(\omega\tau)^4\right\}\right]
\end{equation}

The value which determines the magnetic moment
\begin{equation}
Z={\rm Re}\left[F(x_1)-F(x_2)\right],
\end{equation}
where
\begin{equation}\label{F_2_1}
F(x)=\frac{2J_2(x_2^*)}{(x^2-x_2^{*2})\left[J_0(x_2^*)-J_2(x_2^*)\right]}-\frac{2x_2^{*2}J_2(x)}{x^2(x^2-x_2^{*2})\left[J_0(x)-J_2(x)\right]}
\end{equation}
Expanding the parts of this expression we get:
\begin{equation}
\frac{2J_2(x_2^*)}{J_0(x_2^*)-J_2(x_2^*)}=\frac{1}{4}x_2^{*2}+\frac{7}{96}x_2^{*4}+O\left(x_2^{*6}\right)
\end{equation}
Then taking into account that $x_2^{*2}=-x_2^2$ we find:
\begin{equation}\label{F_x1}
F(x_2)=\frac{7}{96}x_2^2+O(x_2^6),~~~{\rm and ~therefore}~~{\rm Re}\left[F(x_2)\right]=O(x_2^6).
\end{equation}
To find $F(x_1)$ we need to perform a more subtle expansion over $x_1=u_1+iv_1$ since $|u_1|\ll 1$ and $|v_1|\gg 1$. With the exponential accuracy we obtain:
\begin{equation}
\frac{2J_2(x_1)}{x_1^2}=\frac{\sqrt{2}}{\sqrt{\pi}}\left\{1-iu_1-\frac{1}{2}u_1^2\right\}\frac{1}{v_1^{5/2}}e^{v_1}
\end{equation}
\begin{equation}
J_0(x_1)-J_2(x_1)=\frac{\sqrt{2}}{\sqrt{\pi}}\left\{1-iu_1-\frac{1}{2}u_1^2\right\}\frac{1}{v_1^{1/2}}e^{v_1}
\end{equation}
Then we obtain:
\begin{equation}
Z={\rm Re}\left[F(x_1)\right]=\frac{1}{2}\left(\frac{\xi}{R}\right)^2{\rm Re}\left\{\frac{1}{(1-i\omega\tau)}\left(\frac{1}{4}x_2^2-\frac{7}{96}x_2^4\right)-\frac{x_2^2}{(1-i\omega\tau)}\frac{1}{2}\left(\frac{\xi}{R}\right)^2\right\}
\end{equation}
In the leading order over $(R/\xi)$ we get:
\begin{equation}
Z=\frac{7}{192}(\omega\tau)^2\left(\frac{R}{\xi}\right)^2
\end{equation}

\subsection{The limit $R\gg \xi$ and $(\xi/R)^2\ll\omega\tau\ll (\xi/R)$}

As in the previous section the expansion for the value $x_1$ reads
\begin{equation}
x_1=\frac{(\omega\tau)}{2\sqrt{2}}\frac{R}{\xi}\left[1+O\left\{(\omega\tau)^2\right\}\right]+\sqrt{2}i\frac{R}{\xi}\left[1+\frac{(\omega\tau)^2}{32}+O\left\{(\omega\tau)^4\right\}\right]
\end{equation}
However, now the value $x_2$ becomes large. Then to calculate $F(x_2)$ we can use Eq.~(\ref{F_5}):
\begin{equation}\label{F_5_g}
{\rm Re}\left[F(x_2)\right]={\rm Re}\left\{-\frac{\left(1-\frac{15}{8}\frac{i}{x_2}-\frac{105}{128}\frac{1}{x_2^2}\right)}{2x_2^2\left(1-\frac{7}{8}\frac{i}{x_2}-\frac{57}{128}\frac{1}{x_2^{2}}\right)}-\frac{\left(1-\frac{15}{8}\frac{1}{x_2}+\frac{105}{128}\frac{1}{x_2^{2}}\right)}{2x_2^2\left(1-\frac{7}{8}\frac{1}{x_2}+\frac{57}{128}\frac{1}{x_2^{2}}\right)}\right\}
\end{equation}
After simplifications we get:
\begin{equation}\label{F_5_g_2}
{\rm Re}\left[F(x_2)\right]={\rm Re}\left\{-\frac{1}{2x_2^2}\left[2-\frac{(1+i)}{x_2}+O\left(\frac{1}{x_2^3}\right)\right]\right\}=O\left(\frac{1}{x_2^5}\right)
\end{equation}
For the function $F(x_1)$ with the exponential accuracy in the leading order we find:
\begin{equation}\label{Fx1_1}
{\rm Re}\left[F(x_1)\right]={\rm Re}\left\{-\frac{\left(1-\frac{15}{8}\frac{1}{x_2}+\frac{105}{128}\frac{1}{x_2^{2}}\right)}{(x_1^2+x_2^{2})\left(1-\frac{7}{8}\frac{1}{x_2}+\frac{57}{128}\frac{1}{x_2^{2}}\right)}+\frac{x_2^{2}}{v_1^2(x_1^2+x_2^{2})}\right\}
\end{equation}
\begin{equation}\label{Fx1_2}
Z=\frac{1}{2}\left(\frac{\xi}{R}\right)^2
\end{equation}

\subsection{The limit $R\gg \xi$ and $(\xi/R)\ll\omega\tau\ll 1$}

In this limit, both $x_1$ and $x_2$ are large and ${\rm Re}\left[F(x_2)\right]$ is negligibly small. Then using Eq.~(\ref{Q_3}) we find:
\begin{equation}\label{Q_g_3}
Z=\frac{1}{2}\left(\frac{\xi}{R}\right)^2
\end{equation}

\end{widetext}

\end{document}